\documentclass[sigconf]{acmart}
\usepackage[ruled,vlined]{algorithm2e}
\usepackage{subfigure}
\usepackage{array}
\usepackage{graphicx}
\usepackage{multirow}
\usepackage{bm}
\usepackage{enumitem}
\usepackage{setspace}
\usepackage{tcolorbox}
\newcommand{\tabincell}[2]{\begin{tabular}{@{}#1@{}}#2\end{tabular}}
\AtBeginDocument{%
  \providecommand\BibTeX{{%
    \normalfont B\kern-0.5em{\scshape i\kern-0.25em b}\kern-0.8em\TeX}}}

\makeatletter
\newcommand\footnoteref[1]{\protected@xdef\@thefnmark{\ref{#1}}\@footnotemark}
\makeatother

\author{Shangbin Feng}
\affiliation{%
 \institution{Xi'an Jiaotong University}
 \city{Xi'an}
\country{China}}
\email{wind_binteng@stu.xjtu.edu.cn}

\author{Herun Wan}
\affiliation{%
  \institution{Xi'an Jiaotong University}
  \city{Xi'an}
  \country{China}}
\email{wanherun@stu.xjtu.edu.cn}

\author{Ningnan Wang}
\affiliation{%
  \institution{Xi'an Jiaotong University}
  \city{Xi'an}
  \country{China}}
\email{mrwangyou@stu.xjtu.edu.cn}

\author{Jundong Li}
\affiliation{
  \institution{University of Virginia}
  \city{Charlottesville}
  \country{USA}}
\email{jundong@virginia.edu}

\author{Minnan Luo}
\affiliation{
  \institution{Xi'an Jiaotong University}
  \city{Xi'an}
  \country{China}}
\email{minnluo@xjtu.edu.cn}

\begin{document}
\fancyhead{}


\settopmatter{printacmref=false} 
\renewcommand\footnotetextcopyrightpermission[1]{} 
\pagestyle{plain} 

\title{TwiBot-20: A Comprehensive Twitter Bot Detection Benchmark}

\begin{abstract}
Twitter has become a vital social media platform while an ample amount of malicious Twitter bots exist and induce undesirable social effects. Successful Twitter bot detection proposals are generally supervised, which rely heavily on large-scale datasets. However, existing benchmarks generally suffer from low levels of user diversity, limited user information and data scarcity. Therefore, these datasets are not sufficient to train and stably benchmark bot detection measures. To alleviate these problems, we present TwiBot-20, a massive Twitter bot detection benchmark, which contains 229,573 users, 33,488,192 tweets, 8,723,736 user property items and 455,958 follow relationships.  TwiBot-20 covers diversified bots and genuine users to better represent the real-world Twittersphere. TwiBot-20 also includes three modals of user information to support both binary classification of single users and community-aware approaches.  To the best of our knowledge, TwiBot-20 is the largest Twitter bot detection benchmark to date. We reproduce competitive bot detection methods and conduct a thorough evaluation on TwiBot-20 and two other public datasets. Experiment results demonstrate that existing bot detection measures fail to match their previously claimed performance on TwiBot-20, which suggests that Twitter bot detection remains a challenging task and requires further research efforts.
\end{abstract}


\maketitle

\section{Introduction}
\label{sec:introduction}
Twitter is one of the most popular social media platforms and as reported by Statista\footnote{\url{https://www.statista.com/}}, where millions of daily active Twitter users using the platform. Twitter is also free to use and easy to access, which encourages individuals as well as organizations to view and publish contents of interests. Besides being used by genuine users, Twitter is home to an ample amount of automated programs, which are also known as Twitter bots. Some Twitter bots exploit Twitter features to pursue malicious goals such as election interference~\cite{10.1145/3308560.3316486, DBLP:journals/corr/Ferrara17aa} and extreme propaganda~\cite{berger2015isis}. Twitter bots co-exist with human users on Twitter and they hide their automated nature by imitating genuine users. Since identifying bots in social media is crucial to preserving the integrity of the online discourse and these proposals are generally supervised, many research efforts have been devoted to the creation of relevant datasets.


Through the past decade, many bot detection datasets have been proposed and used. Regarding the user composition of datasets, the pronbots~\cite{yang2019arming} dataset contains only Twitter bots where bot detection is treated as a task of outlier detection, while a majority of datasets, such as varol-icwsm~\cite{varol-icwsm} and cresci-17~\cite{cresci2017paradigm}, contains both human and bot users to form a binary classification problem. Regarding user information, while caverlee~\cite{caverlee} provides semantic and property information of Twitter users, gilani-17~\cite{gilani-17} and others only include property information, and they vary greatly in user information completeness. Regarding dataset size, the number of users included in each dataset also range from 693 users in cresci-rtbust~\cite{cresci-rtbust} to 50,538 in midterm-18~\cite{yang2020scalable}, and one of the most widely used dataset cresci-17~\cite{cresci2017paradigm} contains 2,764 human users and 7,049 bots.

While previous research efforts have provided an ample amount of bot detection datasets for training and evaluation, they generally suffer from the following issues and fail to present a stable benchmark:
\begin{itemize}[topsep=4pt, leftmargin=*]
    \item \textbf{User diversity.} Existing bot detection datasets often focus on a specific type or cluster of users, failing to capture diversified bots that co-exist on the real-world Twittersphere. For example, dataset midterm-18~\cite{yang2020scalable} only contains users that are actively involved in the 2018 US midterm elections. As a result, it fails to evaluate existing methods' ability to identify bots outside politics.
    \item \textbf{Limited user information.} Twitter users possess semantic, property and neighborhood information, while existing benchmarks only include a small fraction of multi-modal user information and fall short of comprehensiveness. For instance, the widely adopted cresci-17~\cite{cresci2017paradigm} only contains semantic and property information, failing to incorporate user neighborhood information to support community-based bot detection approaches.
    \item \textbf{Data scarcity.} Previous small-scale datasets are not sufficient to train and stably benchmark novel bot detection measures, hindering the development of new approaches. To the best of our knowledge, the largest existing bot detection dataset midterm-18~\cite{yang2020scalable} contains 50,538 users, leaving increasingly complex bot detectors data-hungry.
\end{itemize}

In light of the drawbacks of previous bot detection datasets, we collect and annotate a comprehensive Twitter bot detection benchmark TwiBot-20, which can alleviate the lack of user diversity, limited user information and data scarcity problems:
\begin{itemize}[leftmargin=*]
    \item We conduct BFS-based traversal through follow relationships, starting from a large number of seed users in different topics. As a result, users in TwiBot-20 are diverse in geographical locations and interest domains, making TwiBot-20 more representative of the current Twittersphere.
    \item We retrieve all three modals of user information, namely semantic, property and neighborhood information from Twitter API to facilitate leveraging multi-modal user information. To the best of our knowledge, TwiBot-20 is the first publicly available bot detection dataset that includes user follow relationships.
    \item To the best of our knowledge, TwiBot-20 establishes the largest benchmark of Twitter bot detection to date, which contains 229,573 Twitter users, 8,723,736 user property items, 33,488,192 tweets and 455,958 follow links.
\end{itemize}


In the following, we first review related work in Section ~\ref{sec:related}, then define the task of Twitter bot detection in Section ~\ref{sec:problemdefinition}. We detail the collection and annotation process of TwiBot-20 in Section ~\ref{sec:datacollection} and conduct in-depth data analysis in Section ~\ref{sec:dataanalysis}. We conduct extensive experiments to evaluate existing bot detection measures on TwiBot-20 in Section ~\ref{sec:experiment}, and conclude the paper in Section ~\ref{sec:conclusion}.

\section{Related Work}
\label{sec:related}
In this section, we briefly review the related literature in social media bot detection and Twitter bot detection datasets.

\subsection{Social Media Bot Detection}
The first generation of proposals for Twitter bot detection focuses on feature engineering with user information. Lee \textit{et al.} \cite{lee2013warningbird} proposed to verify URL redirections in tweets. Thomas \textit{et al.} \cite{thomas2011design} focused on classifying mentioned websites. Gao \textit{et al.}~\cite{gao2012towards} merges spam tweets into campaigns for bot detection. Yang \textit{et al.}~\cite{yang2013empirical} designed novel features to counter the evolution of Twitter bots. Other features are also adopted such as information on the user homepage ~\cite{lee2011seven}, social networks ~\cite{minnich2017botwalk}, and timeline of accounts ~\cite{cresci2016dna}.

With the advent of deep learning, neural networks are increasingly adopted for Twitter bot detection. 
Stanton \textit{et al.}~\cite{stanton2019gans} leveraged generative adversarial networks for Twitter bot detection.
Wei \textit{et al.}~\cite{wei2019twitter} used recurrent neural networks to identify bots with tweet semantics.
Kudugunta \textit{et al.}~\cite{kudugunta2018deep} jointly leveraged user features and tweet semantics to propose a LSTM-based bot detector.
Alhosseini \textit{et al.}~\cite{ali2019detect} proposed to adopt graph convolutional networks to leverage user features and the Twittersphere structure.



\subsection{Twitter Bot Detection Datasets}
An ample amount of datasets are proposed to train and evaluate bot detection methods. One of the earliest bot detection dataset is caverlee-2011~\cite{lee2011seven}, which is collected from December 30, 2009 to August 2, 2010 on Twitter with 22,223 content polluters and 19,276 legitimate users. Another early dataset is cresci-2015~\cite{cresci2015fame} which provides a dataset of genuine accounts and fake followers.

An increasing amount of bot detection benchmarks are proposed since 2017. varol-2017~\cite{varol2017online} contains manual annotation of 2,573 Twitter accounts. vendor-purchased-2019~\cite{yang2019arming} consists of fake follower accounts purchased online. To the best of our knowledge, the largest Twitter bot detection dataset to date is midterm-18~\cite{yang2020scalable}, providing 50,538 annotated users. Apart from that, verified-2019~\cite{yang2020scalable}, botwiki-2019~\cite{yang2020scalable}, cresci-rtbust-2019~\cite{mazza2019rtbust}, Astroturf~\cite{sayyadiharikandeh2020detection} are also more recent bot detection datasets of various size and information completeness.

\begin{algorithm}[h]
    \caption{TwiBot-20 User Selection Strategy}
    \label{alg:ScfT}
    \KwIn{initial seed user $u_0$ in a user cluster}
    \KwOut{user information set $F$}
    \BlankLine
    $u_0.layer\ \ \overleftarrow{} \ \ $0; // designate seed user as layer 0\\
    $S\ \ \overleftarrow{}\ \ \{u_0\}$; // set of users to expand\\
    $u_0.expanded\ \ \overleftarrow{}\ \ False$;\\
    $F \ \overleftarrow{}\ \ \varnothing$; \\
    \While{$S \neq \varnothing$}{
        $u\ \ \overleftarrow{}\ \ S.pop()$; // expand with user $u$ \\
        $T(u)\ \ \overleftarrow{}\ \ get\_tweet(u)$;\\
        $P(u)\ \ \overleftarrow{}\ \ get\_property(u)$;\\
        \If{$u.layer \geq 3\ $\textbf{or}$\ u.expanded == True$}{
            $F\ \ \overleftarrow{}\ \ F \cup u(T,P,N = \varnothing)$;\\
            \textbf{continue;} // three layers max
        }
        $u.expanded\ \ \overleftarrow{}\ \ True$;\\
        $N^f(u)\ \ \overleftarrow{}\ \ get\_friend(u)$;\\
        $N^t(u)\ \ \overleftarrow{}\ \ get\_follower(u)$;\\
        $N(u)\ \ \overleftarrow{}\ \ \{N^f(u), N^t(u)\}$;\\
        $F\ \ \overleftarrow{}\ \ F \cup u(T,P,N)$;\\
        $S\ \ \overleftarrow{}\ \ S\cup N^f(u)\cup N^t(u)$;\\
        \For{$y \in N^f(u) \cup N^t(u)$}{
            $y.expanded\ \overleftarrow{}\ False$;\\
            $y.layer\ \overleftarrow{}\ u.layer+1$;\\
        }
    }
    \textbf{Return} $F$; // obtained one cluster of user information
\end{algorithm}

\section{Problem Definition}
\label{sec:problemdefinition}
A Twitter bot is a type of bot software that controls a Twitter account via automated programs and the Twitter API. In this paper, we study the problem of Twitter bot detection to identify Twitter bots that pursue malicious goals since they pose threat to the online discourse.

Let $U$ be a Twitter user, consisting of three aspects of user information: semantic $T$, property $P$ and neighborhood $N$. 
Semantic information of Twitter users are user-generated natural language posts and texts, such as tweets and replies. Property information of Twitter users are numerical and categorical user features such as follower count and whether the user is verified. Neighborhood information of Twitter users are their followers and followings, which form the graph structure of the Twittersphere.
Similar to previous research~\cite{yang2020scalable, kudugunta2018deep}, we treat Twitter bot detection as a binary classification problem, where each user could either be human ($y = 0$) or bot ($y = 1$). 

Formally, we can define the Twitter bot detection task as follows. Given a Twitter user $U$ and its information $T$, $P$ and $N$, learn a bot detection function $f:f(U(T,P,N)) \rightarrow \hat{y}$, such that $\hat{y}$ approximates ground truth $y$ to maximize prediction accuracy.

\section{Data Collection}
\label{sec:datacollection}
In this section, we present how to select Twitter users from the Twittersphere, retrieve multi-modal user information and derive trustworthy annotations to construct the benchmark. TwiBot-20 was collected in this way from July to September 2020.

\subsection{User Selection Strategy}
\label{subsec:userselection}
To diversify user sampling in order to better approximate the current Twittersphere, TwiBot-20 employs breadth-first search starting from different root nodes, which is named seed users in our algorithm. Specifically, we treat users on the Twittersphere as nodes and their follow relationship as edges to form a directed graph. For each seed user, it is placed in layer $0$ of that user cluster. Users in layer $i+1$ are expanded from users in layer $i$ along their follow edges. Such an expansion process ends at layer $3$ and forms a user cluster. TwiBot-20 merges user clusters starting from different seed users to form the complete dataset. TwiBot-20's user selection strategy is presented in Algorithm ~\ref{alg:ScfT}, following notations defined in Section ~\ref{sec:problemdefinition}.

TwiBot-20's user selection strategy is different from previous benchmarks in that it does not demand selected users to follow any given pattern or restrict them to any specific topic. Such a relaxation of constraints is crucial for TwiBot-20 to better represent the diversified Twittersphere and evaluate bot detectors' ability to capture multiple types of bots that co-exist on online social media.

\subsection{Seed User Selection}
\label{subsec:seeduser}
As detailed in Section ~\ref{subsec:userselection}, TwiBot-20 is obtained by controlled breadth-first search expanded from different seed users. 
The goal of TwiBot-20 is to accurately represent the diversified Twittersphere to benchmark generalizable bot detection, thus the search algorithm should be designed to cover diverse user groups and communities.
Politics, business, entertainment and sports are four interest domains that would overlap with everyone's online engagements. Thus, unlike previous bot detection datasets which are limited to a specific topic or hashtag, TwiBot-20 selects diverse seed users from these domains to result in a representative collection of the current Twittersphere. Specifically, 40 seed users come from these four distinct disciplines:
\begin{itemize}[leftmargin=*]
    \item Politics: We select national and local politicians from diverse ideological spectrum, major media outlets and influential political commentators as seed users. e.g. {\verb|@SpeakerPelosi|}
    \item Business: We select corporations, financial media outlets and influential business pundits as seed users. e.g. {\verb|@amazon|}
    \item Entertainment: We select well-known artists, comedians and video game streamers as seed users. e.g. {\verb|@samsmith|}
    \item Sports: We select athletes, sports team and sports news outlets from various types of sports as seed users. e.g. {\verb|@StephenCurry30|}
\end{itemize}

In addition to the big names in each area, we also sample users who comment under relevant tweets and active users in relevant hashtags as seed users to provide another view of the community and ensure exhaustiveness. A complete list of all seed users is available in the full dataset TwiBot-20. By using a large number of seed users from different interest domains, we ensure that TwiBot-20 covers diversified users to better represent the current Twittersphere.

\subsection{User Information Selection}
\label{subsec:userinformation}
After determining the user list of TwiBot-20, we use Twitter API to retrieve all three aspects of user information as defined in Section ~\ref{sec:problemdefinition}. Specifically,
\begin{itemize}[leftmargin=*]
    \item For semantic information, we retrieve the most recent 200 tweets of each Twitter user to capture its recent activities. We preserve the original form of the multilingual tweet content, leaving emojis, urls and retweets intact for bot detectors to process tweet text in ways they see fit.
    \item For property information, we incorporate all property items provided by the Twitter API for each user. As a result, each user has 38 property items recorded in TwiBot-20.
    \item For neighborhood information, we retrieve followers and followings of a Twitter user and record follow relationships between users in TwiBot-20.
\end{itemize}

Existing bot detection datasets mostly leave out neighborhood information and contains only a fraction of property items or user tweets. In comparison, TwiBot-20 includes all user information that is directly retrievable from the Twitter API so that future bot detectors could leverage whatever user information they see fit without being constrained to the scope of the dataset.

\subsection{Data Annotation Strategy}
\label{subsec:annotationstrategy}
Data annotation in Twitter bot detection is particularly difficult and prone to bias, thus we employ a specialized data annotation strategy that draws on conclusions of previous research efforts and emphasize trustworthiness. Firstly, we summarize previous literature and propose general criteria to identify a bot user, which are listed as follows:
\begin{itemize}[leftmargin=*]
    \item Lack of pertinence and originality in tweets. 
    \item Highly automated activity and API usage. 
    \item Tweets containing external link promoting phishing or commercials.
    \item Repeated tweets with identical content. 
    \item Tweets containing irrelevant URLs. 
\end{itemize}

Guided by these standards, we launch a crowdsourcing campaign at Appen\footnote{\url{https://www.appen.com/}}. According to our contract, annotators should be active Twitter users and are required to read a guidelines document in which we explain the five characteristics of bots along with representative examples. Five annotators are then assigned to each user in TwiBot-20 to determine whether it is operated by bot or not. In order to identify potentially ambiguous cases, annotators are permitted to report 
‘undecided’ regarding a specific user. We also designed standard questions where the user is clearly a bot or human. We mix these standard questions with annotation inquiries to evaluate an annotator's performance. Annotators who are more than 80\% correct on standard questions are considered to be trustworthy and their annotation is adopted. As a result, the crowdsourcing campaign provided 63,816 annotation records and takes approximately 33 days.

\begin{table}
 \caption{Statistics of different bot detection benchmarks, from left to right, user count, user property item count, total tweet count and follow relationship count in each dataset. }
 \label{tab:DatasetStatistics}
 \begin{tabular}{c c c c c} 
 \toprule
   Dataset & \#User & \#Property & \#Tweet & \#Follow \\ \hline

  varol-icwsm~\cite{varol-icwsm} & 2,573 & 0 & 0 & 0 \\
  
  pronbots~\cite{yang2019arming} & 21,965 & 750,991 & 0 & 0 \\
  
  celebrity~\cite{yang2019arming} & 5,971 & 879,954 & 0 & 0 \\
  
  gilani-17~\cite{gilani-17} & 2,653 & 104,515 & 0 & 0 \\
  
  cresci-rtbust~\cite{cresci-rtbust} & 693 & 28,968 & 0 & 0 \\
  
  cresci-stock~\cite{Cresci-stock} & 13,276 & 551,603 & 0 & 0 \\
  
  midterm-18~\cite{yang2020scalable} & 50,538 & 909,684 & 0 & 0 \\
  
  botwiki~\cite{yang2020scalable} & 698 & 29,082 & 0 & 0 \\
  
  verified~\cite{yang2020scalable} & 1,987 & 83,383 & 0 & 0 \\
  
  PAN-19\footnoteref{PAN-19}& 11,568 & 0 & 369,246 & 0 \\
  
  caverlee~\cite{caverlee} & 22,224 & 155,568 & 5,613,166 & 0  \\
  
  cresci-17~\cite{cresci2017paradigm} & 14,398 & 547,124 & 18,179,186 & 0 \\ \hline
  
  \textbf{TwiBot-20} & \textbf{229,573} & \textbf{8,723,736} & \textbf{33,488,192} & \textbf{455,958} \\
 
 \bottomrule
 \end{tabular}
 \end{table}

Although we provided annotation guidelines, assigned five annotators to each user and designed standard questions for performance evaluation, the crowdsourcing results are not reliable on its own. We further take the following steps to reach the final annotation of users in TwiBot-20:
\begin{itemize}[leftmargin=*]
    \item Firstly, if a user is verified by Twitter, we consider it to be a genuine user.
    \item For remaining users, if four out of five annotators believe that it is bot or human, we annotate the user accordingly.
    \item For other users with less mutual agreement from crowdsourcing, we use Twitter's direct message feature to send out simple questions in natural language, collect answers from users that respond and manually determine their annotations.
    \item Finally, remaining undecided users are manually examined within our research team. To ensure the trustworthiness of these ambiguous users, we discard disputed cases and only annotate when we reach a consensus on a Twitter user. 
\end{itemize}

Data annotation in Twitter bot detection is heavily influenced by underlying assumptions and subject to bias. We synthesize previous literature, launch a carefully designed crowdsourcing campaign and follow a meticulous process to reach the final annotation of a specific user. Further analysis in Section ~\ref{subsec:annotationquality} would further demonstrate the trustworthiness of TwiBot-20 annotations.

\begin{table}
 \caption{User information modalities that each bot detection dataset contains. The definition of semantic, property and neighborhood information follows that of Section 
 ~\ref{sec:problemdefinition}. varol-icwsm~\cite{varol-icwsm} is considered to not contain any aspect of user information since it is merely a list of user ids that are considered to be bots or human.}
 \label{tab:SPN}
 \begin{tabular}{c c c c} 
 \toprule
   Dataset & Property & Semantic & Neighbor \\ \hline
   
  varol-icwsm~\cite{varol-icwsm} & & & \\
  
  pronbots~\cite{yang2019arming} & \checkmark & & \\
  
  celebrity~\cite{yang2019arming} & \checkmark & & \\
  
  gilani-17~\cite{gilani-17} & \checkmark & & \\
  
  cresci-rtbust~\cite{cresci-rtbust} & \checkmark & & \\
  
  cresci-stock~\cite{Cresci-stock} & \checkmark & & \\
  
  midterm-18~\cite{yang2020scalable} & \checkmark & & \\
  
  botwiki~\cite{yang2020scalable} & \checkmark & & \\
  
  verified~\cite{yang2020scalable} & \checkmark & & \\
  
  PAN-19\footnoteref{PAN-19} & & \checkmark & \\
  
  caverlee~\cite{caverlee} & \checkmark & \checkmark \\
   
  cresci-17~\cite{cresci2017paradigm} & \checkmark & \checkmark & \\ \hline
  
  \textbf{TwiBot-20} & \checkmark & \checkmark & \textbf{\checkmark} \\
 
 \bottomrule
 \end{tabular}
 \end{table}

\subsection{Data Release}
Users in TwiBot-20 are determined in Section ~\ref{subsec:userselection} and Section ~\ref{subsec:seeduser}. Multi-modal user information is retrieved from Twitter API as described in Section ~\ref{subsec:userinformation} and Section ~\ref{subsec:annotationstrategy} details how we created trustworthy annotations for TwiBot-20. We release these data according to the following process:

\begin{itemize}[leftmargin=*]
    \item We conduct a random partition of 7:2:1 on the labeled users to obtain the train, validation and test set of the TwiBot-20 benchmark. In order to preserve the dense graph structure follow relationship forms, we also provide unsupervised users as the support set of TwiBot-20.
    \item We organize users in each set into the JSON format to obtain four data files: {\verb|train.json|}, {\verb|dev.json|}, {\verb|test.json|} and {\verb|support.json|}. For each user, we provide user IDs to identify users and all their semantic, property and neighborhood information collected in Section ~\ref{subsec:userinformation}.
    \item We release TwiBot-20 with succinct documentation at \textbf{\url{https://github.com/BunsenFeng/TwiBot-20}}. A small sample of TwiBot-20 is directly available at the github repository, while researchers are also encouraged to download and use the full TwiBot-20 to facilitate bot detection research.
\end{itemize}

\section{Data Analysis}
\label{sec:dataanalysis}
In this section, we firstly contrast the size of different bot detection datasets. We then compare TwiBot-20's information completeness and user diversity with other benchmarks. Finally we conduct annotation analysis to demonstrate the trustworthiness of TwiBot-20 annotations.

\subsection{Dataset Size Analysis}
\label{subsec:datasetsize}
Bot detection on social media focuses on solving the real-world problem of online bot presence. In order to well represent real-world genuine users and Twitter bots, a bot detection dataset should be considerable in size to achieve such purpose. We contrast TwiBot-20 and major bot detection datasets regarding dataset size in Table ~\ref{tab:DatasetStatistics}.

\begin{figure*}[h]
  \centering
  \includegraphics[width=\linewidth]{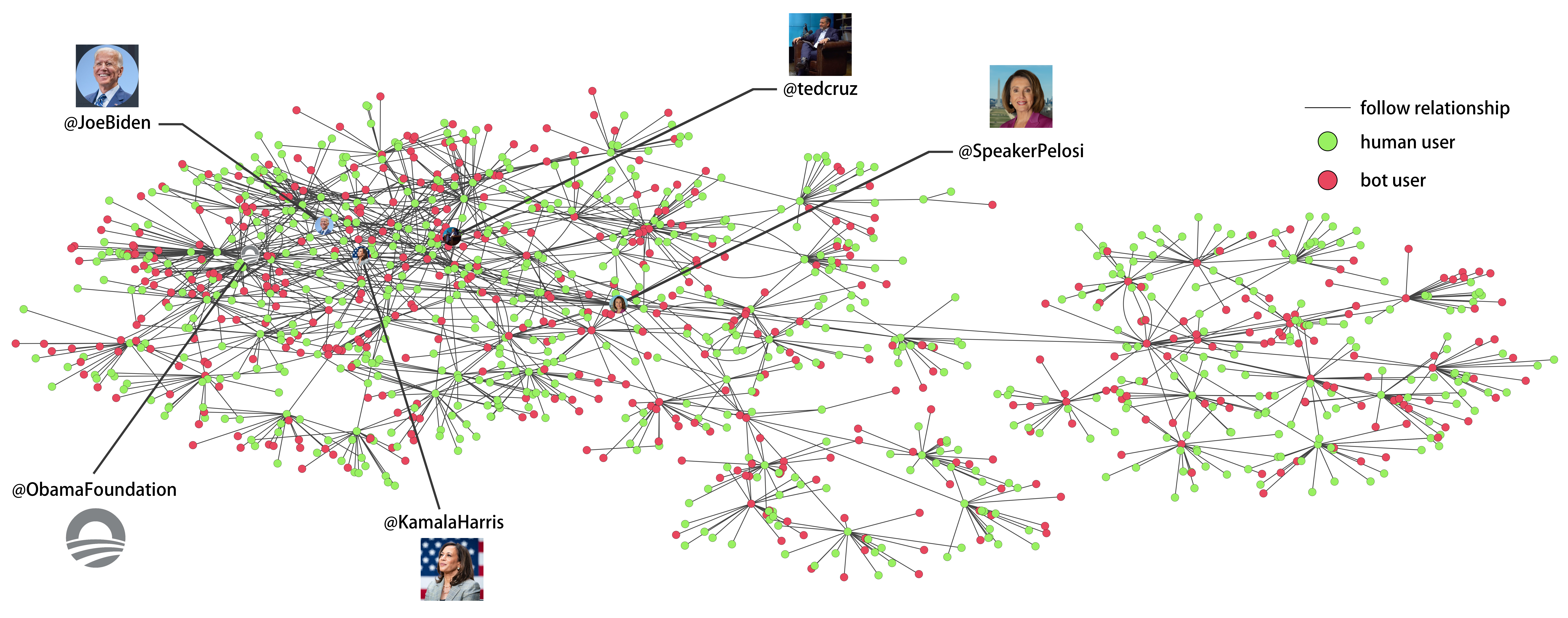}
  \caption{Illustration of a user cluster in TwiBot-20 with \textit{@SpeakerPelosi} as the seed user. Green nodes represent human users in the user cluster, red nodes represent bot users and edges in the graph indicates that one user follows the other.}
  \Description{user cluster illustration}
  \label{fig:pelosi}
\end{figure*}


Table ~\ref{tab:DatasetStatistics} demonstrates that TwiBot-20 leads with a total of 229,573 users, significantly outnumbering previous bot detection datasets. TwiBot-20 also provides a support set that includes massive amount of unsupervised users. This support set enables novel trends such as semi-supervised learning to merge with bot detection research. To the best of our knowledge, TwiBot-20 establishes the largest bot detection benchmark to date and is the first to provide unsupervised users in a bot detection dataset.

\subsection{User Information Analysis}
Online social media is becoming increasingly complex, with users generating huge volumes of multi-modal data every day. Twitter bots are also leveraging this information complexity to evade previous bot detection measures. Twitter users often generate large volumes of multi-modal data, thus bot detection datasets should incorporate all three modals of user information to allow comprehensive analysis of user's behaviour, which might boost bot detection performance and robustness. We study major bot detection datasets and depict their user information completeness in Table ~\ref{tab:SPN}.

Table ~\ref{tab:SPN} demonstates that TwiBot-20 contains all three aspects of user information. Dataset cresci-17~\cite{cresci2017paradigm} and caverlee~\cite{caverlee} contain both semantic and property information, while all other existing baselines only include user semantic or property information. Further exploration shows that datasets with only property information often leaves out certain property items, introducing inevitable bias in the process. To the best of our knowledge, TwiBot-20 is the first publicly available Twitter bot detection benchmark that incorporates user neighborhood information. TwiBot-20's information completeness enables novel bot detectors to leverage as much user information as it could be explicitly retrieved.

To further explore TwiBot-20's user neighborhood information, we illustrate a clutser of users in TwiBot-20 and their follow relationship in Figure ~\ref{fig:pelosi}. It is demonstrated that follow relationship in TwiBot-20 forms a dense graph structure to enable community-based bot detection measures such as graph neural networks.

\subsection{User Diversity Analysis}
An important aim of TwiBot-20 is to accurately represent the diversified Twittersphere and capture different types of bots that co-exist on social media. We study the distribution of profile locations and user interests to examine whether TwiBot-20 has achieved the goal of user diversity.



\begin{figure}[h]
  \centering
  \includegraphics[width=\linewidth]{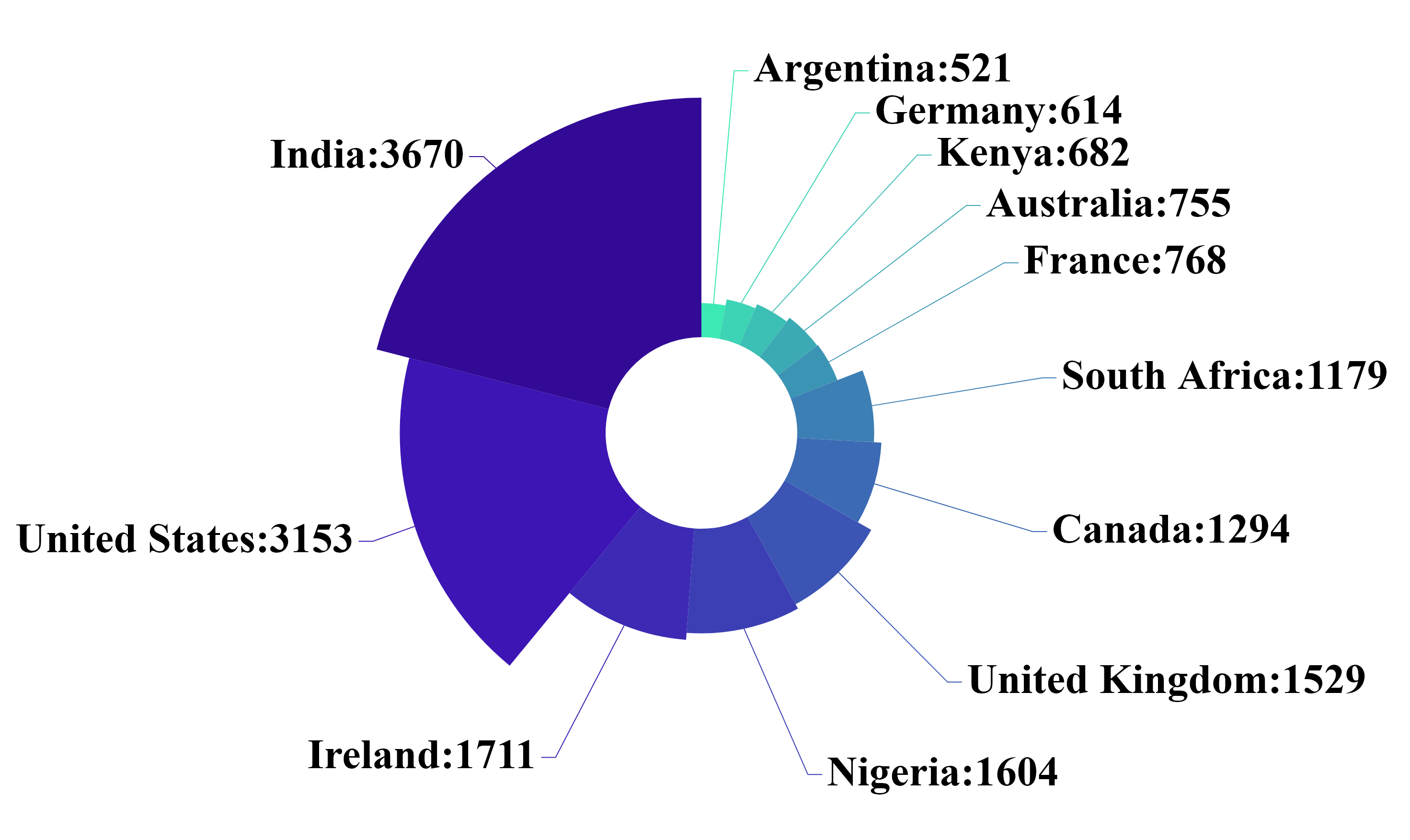}
  \caption{Most frequent countries of user location in TwiBot-20 and their number of appearances. There are 179 countries that appear less than 500 times and they collectively appear 11835 times in TwiBot-20, which are omitted in the figure to preserve clarity.}
  \Description{Here lives some people}
  \label{fig:diversity_location}
\end{figure}

\noindent \textbf{Geographic diversity.} Figure ~\ref{fig:diversity_location} illustrates the geographic location of users in TwiBot-20. While the most frequent two countries are India and the United States, there are also a considerable amount of users from Europe and Africa.

\begin{figure}[h]
  \centering
  \includegraphics[width=\linewidth]{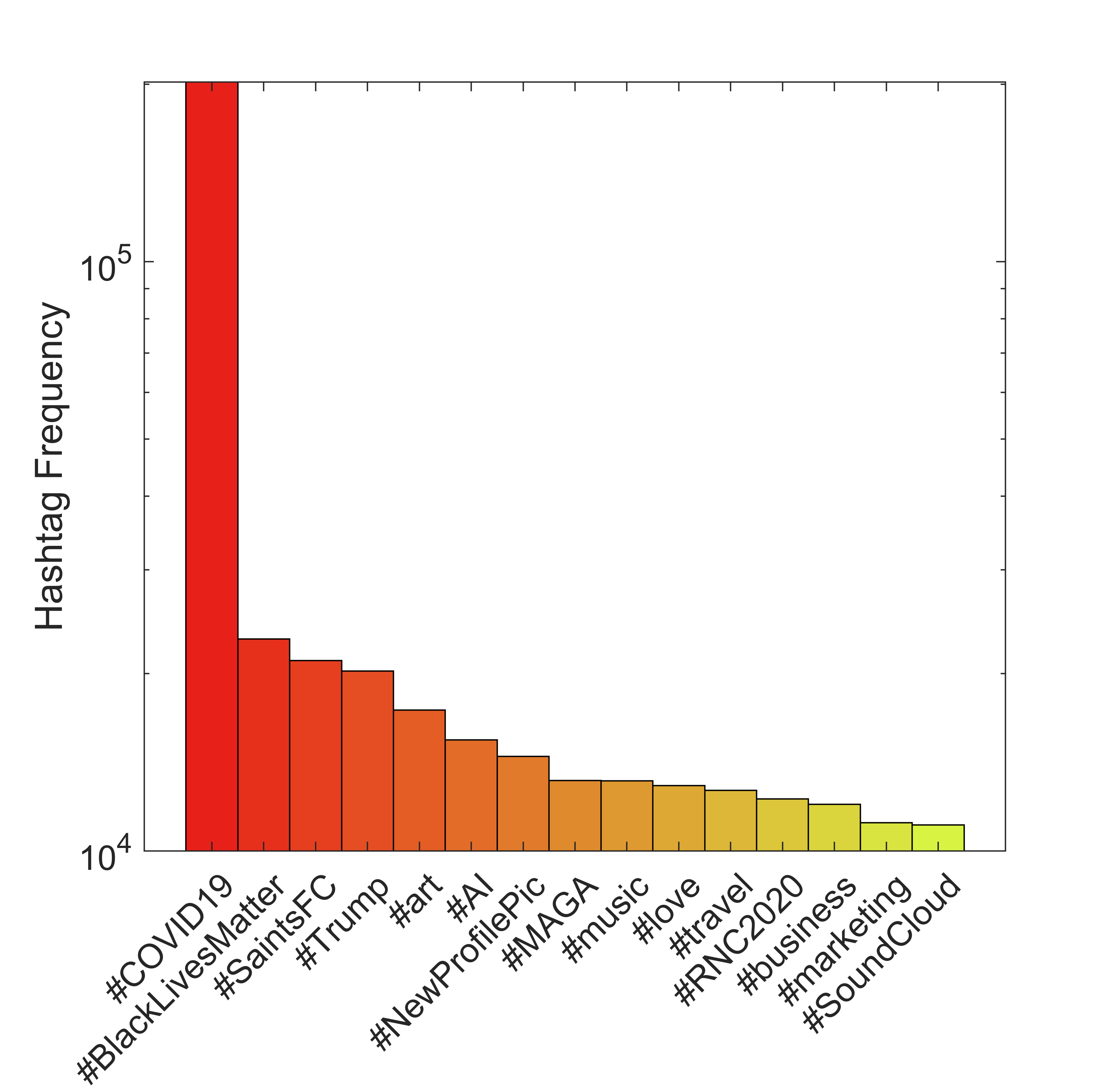}
  \caption{Most frequent hastags in user tweets in our proposed TwiBot-20. We merge similar hashtags such as \#COVID19 and \#coronavirus and select the 15 hashtags of highest frequency.}
  \Description{Here lives some people}
  \label{fig:hashtags}
\end{figure}

\noindent \textbf{User interest diversity.} Figure ~\ref{fig:hashtags} illustrates the most frequently mentioned hashtags and their frequency. The \#COVID19 hashtag ranks first due to the global health crisis at the time of data collection. It is demonstrated that TwiBot-20 captures political users that often tweet with \#Trump and \#RNC2020, sports lovers with \#SaintsFC, business people with \#business and \#marketing, as well as ordinary users that tweet with \#love and \#travel.



Twitter users in TwiBot-20 are thus proved to be diversified in geographic locations and interest domains. TwiBot-20 contains diversified users to better represent the diversified Twittersphere rather than being toy examples of specific scenarios.

\subsection{Annotation Quality Analysis}
\label{subsec:annotationquality}
To prove that our annotation procedure leads to high quality annotation, we examine whether the annotation results are consistent with bot characteristics proposed in previous literature. 

\noindent \textbf{Account Reputation.} We explore the difference between the reputation score~\cite{chu2012detecting} of bot and genuine users in TwiBot-20. Reputation is a coefficient that measures a user's follower count and friend count, defined as follows:
\begin{equation}
    Reputation(u) = \frac{|N^t(u)|}{|N^t(u)|+|N^f(u)|}
\end{equation}

\noindent where $|\cdot|$ denotes the cardinality of a set, $N^t(u)$ denotes the follower set of user $u$ and $N^f(u)$ denotes the following set of user $u$. 

\begin{figure*}
    \centering
    \subfigure[Reputation Score Distribution]{\includegraphics[width=0.32\textwidth]{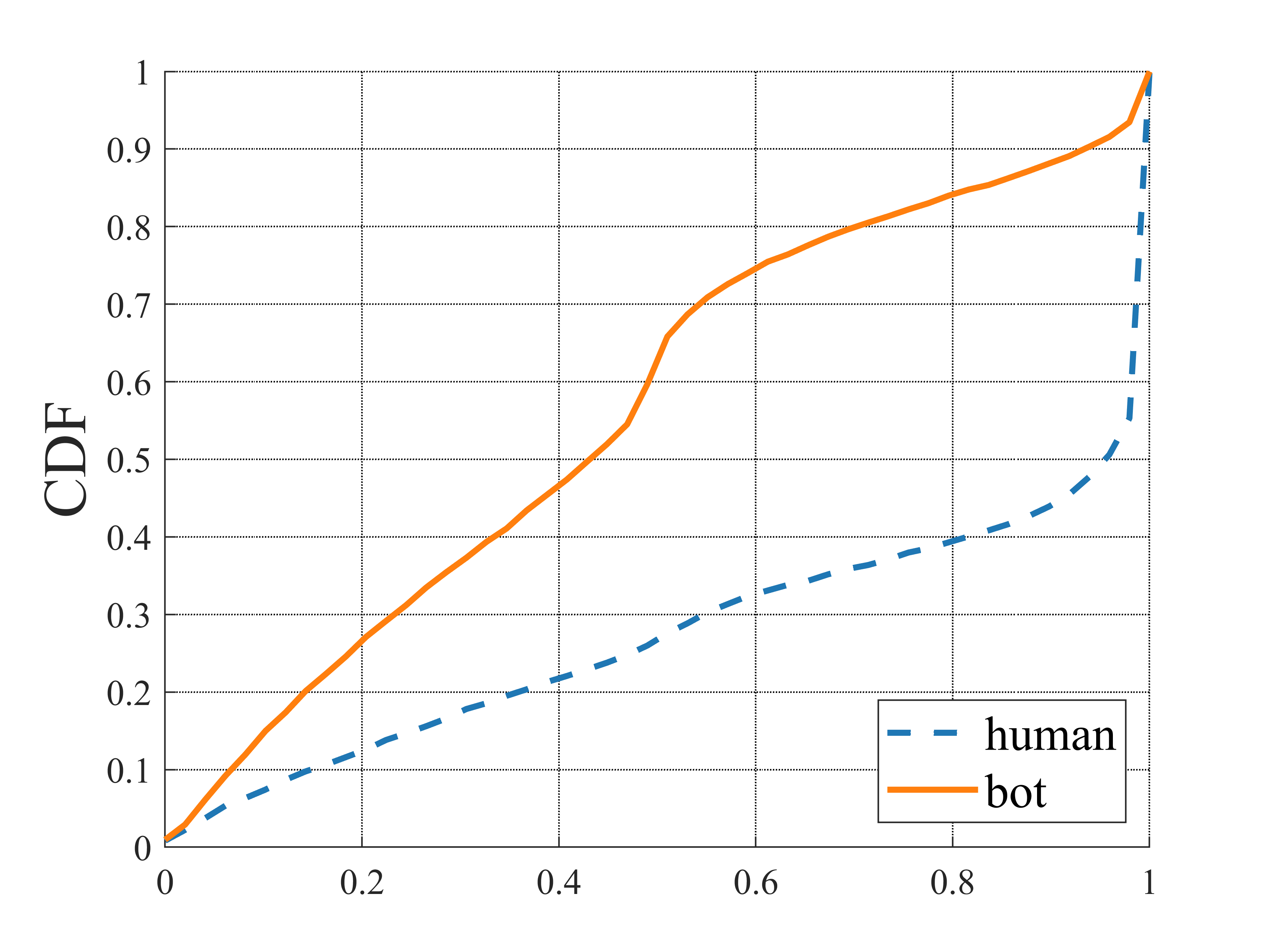}}
    \subfigure[Tweet Count Distribution]{\includegraphics[width=0.32\textwidth]{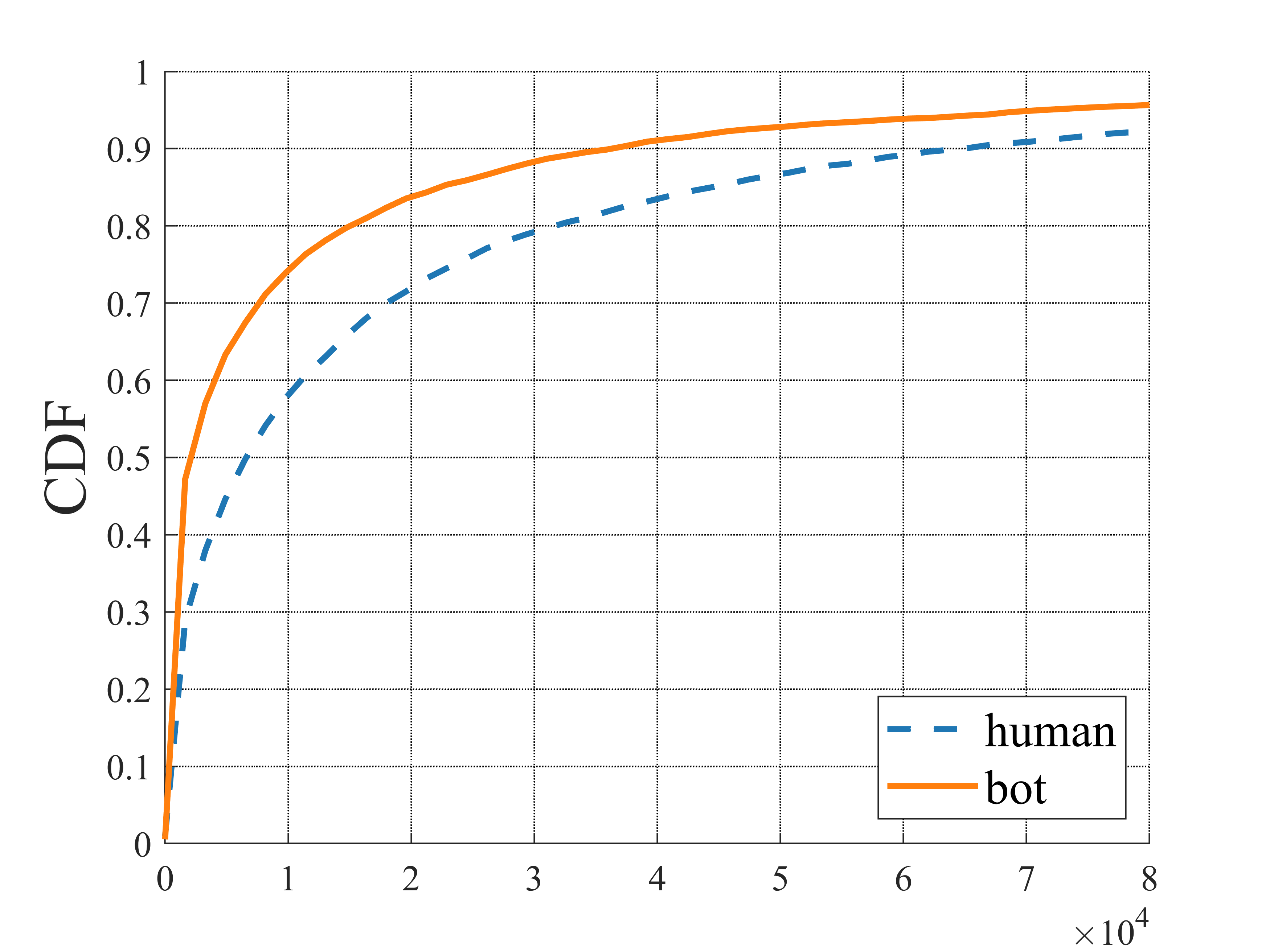}}
    \subfigure[Screen Name Likelihood Distribution]{\includegraphics[width=0.32\textwidth]{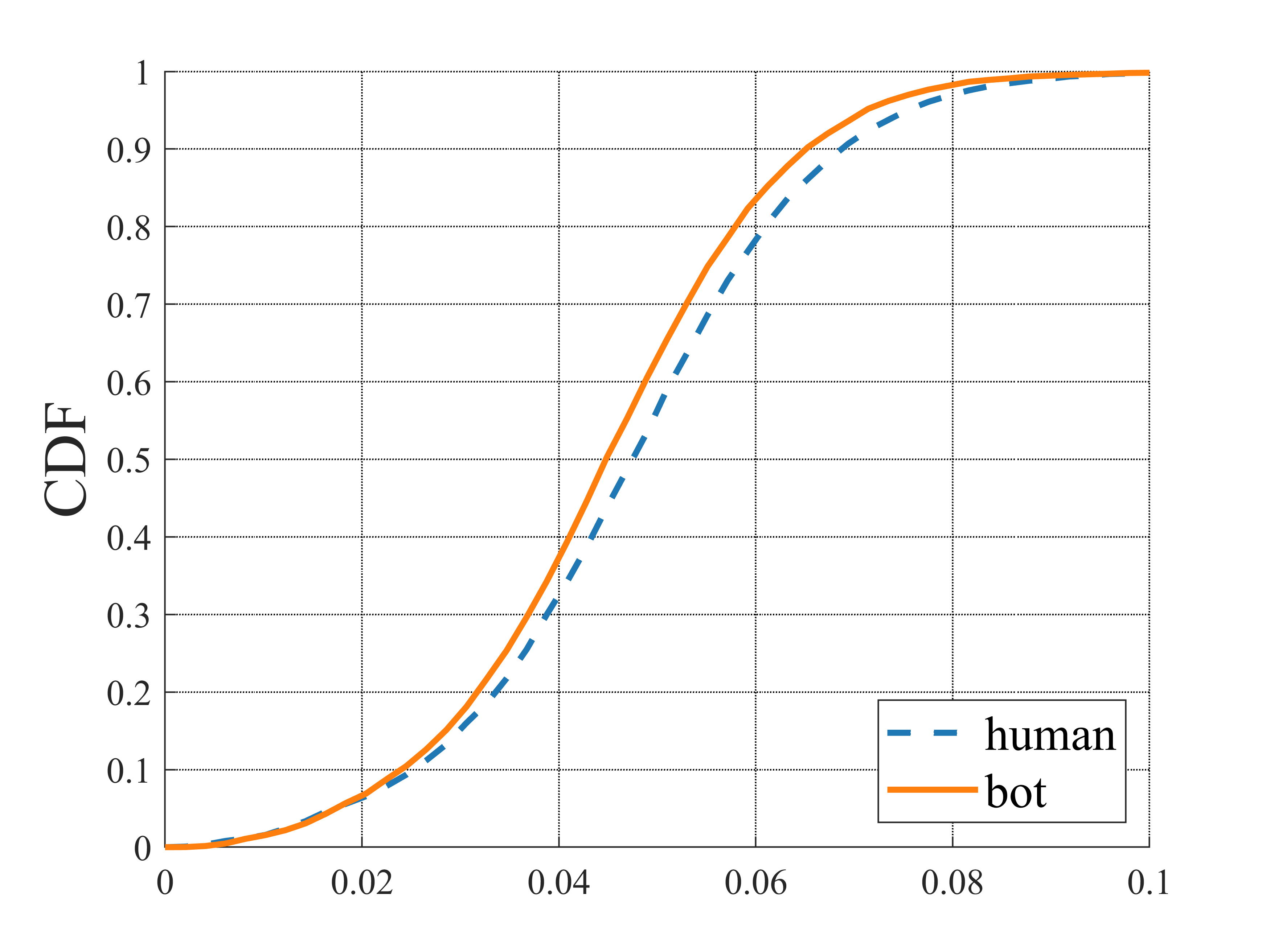}}
    \caption{Cumulative distribution functions of reputation, tweet count and name likelihood among bots and genuine users.}
    \label{fig:CDF}
\end{figure*}

Chu \textit{et al.}~\cite{chu2012detecting} observed that, human users are more likely to follow "famous" or "reputable" users. A celebrity usually has many followers and few friends, and his reputation is close to 1. In contrast, for a bot with few followers and many friends, its reputation is close to 0. We illustrate the cumulative distribution function (CDF) of account reputation for bot and human users in Figure ~\ref{fig:CDF}(a). It is illustrated that genuine users in TwiBot-20 exhibit relatively higher reputation score than bots. Around 60 percent of bots in TwiBot-20 have fewer followers than friends, causing their reputation to be less than 0.5. The reputation score of users in TwiBot-20 matches the proposal of Chu \textit{et al.}~\cite{chu2012detecting}, which strengthens the claim that TwiBot-20 annotation is generally trustworthy.

\noindent \textbf{User Tweet Count.} Perdana \textit{et al.}~\cite{perdana2015bot} observed that bot users would generate a lot of repeated tweets, leading to a higher total tweet count. We explore the relationship between user annotation and its tweet count in TwiBot-20. Figure ~\ref{fig:CDF}(b) shows the CDF of user tweet counts. It is surprising that bot users generate fewer tweets than human. This discrepancy with previous work could be attributed to the fact that bot users have evolved to escape feature-engineering based bot detection and they control the number of tweets now. To further explore this difference, we check the bot users in our dataset and find out the following fact: bot users tweet more frequently in a certain period with long-term hibernation to avoid detection by Twitter, which results in fewer tweets in general. 

\noindent \textbf{Screen Name Likelihood.} Yang \textit{et al.}~\cite{yang2020scalable} observed that bot users sometimes use a random string as their screen name. To measure this feature, they proposed to evaluate the screen name likelihood of users. Twitter only allows letters (upper and lower case), digits and underscores in the screen name field with a 15-character limit. We use the 229,573 screen names in TwiBot-20 and construct the likelihood of all 3,969 possible bigrams. The likelihood of a screen name is defined by the geometric-mean likelihood of all bigrams in it. For a screen name $X$ with length $k$, the likelihood $L(x)$ is defined as follows:

\begin{equation}
    L(x) = [{\prod_{i=1}^{k-1}P(x_i,x_{i+1})}]^{\frac{1}{k-1}}
\end{equation}

\noindent where $x_i$ denotes the i-th character of $X$ and $P(x_i, x_{i+1})$ denotes the likelihood of bigram $(x_i,x_{i+1})$ obtained from the user screen names.

Figure ~\ref{fig:CDF}(c) illustrates the difference between bot users and human users in screen name likelihood. The results indicate that bot users in TwiBot-20 do have slightly lower screen name likelihood, which is compatible with the findings in Yang \textit{et al.}~\cite{yang2020scalable}. We examine some bot users and find that they do have random strings as their screen name such as \textit{@Abolfaz54075615}. Data annotation in TwiBot-20 also matches Yang \textit{et al.}~\cite{yang2020scalable}'s observation, which lends support to TwiBot-20's annotation credibility.

To sum up, TwiBot-20's annotation generally matches previously proposed bot characteristics. 

\section{Experiments}
\label{sec:experiment}
In this section, we conduct extensive experiments and in-depth analysis on TwiBot-20 and two other public datasets to prove the novelty and effectiveness of the proposed benchmark.

\subsection{Experiment Settings}
\label{subsec:expsettings}
\textbf{Datasets.} In addition to our proposed benchmark TwiBot-20, we make use of two other publicly available datasets {\verb|cresci-17|}~\cite{cresci2017paradigm} and {\verb|PAN-19|} \footnote{\label{PAN-19}\url{https://pan.webis.de/clef19/pan19-web/author-profiling.html}} to compare baseline performance on different benchmarks. 

{\verb|cresci-17|} ~\cite{cresci2017paradigm} is partitioned into genuine users, social spam bots, traditional spam bots and fake followers. We utilize {\verb|cresci-17|} as a whole. As of user information, {\verb|cresci-17|} contains semantic and property information of Twitter users. 

{\verb|PAN-19|}\footnoteref{PAN-19} is a dataset of a Bots and Gender Profiling shared task in the PAN workshop at CLEF 2019. It contains user semantic information.

\begin{table}
\setlength{\tabcolsep}{1pt}
\setlength{\abovecaptionskip}{1pt}
\caption{Overview of three adopted bot detection datasets. S, P and N info stands for semantic, property and neighborhood information respectively.}
\label{tab:dataset}
\begin{tabular}{c c c c c c c c}
\toprule

Dataset & \tabincell{c}{User \\ Count} &  \tabincell{c}{Semantic \\ Info} & \tabincell{c}{Property \\ Info} & \tabincell{c}{Neighbor \\ Info} & \tabincell{c}{Release\\ Year} \\ \midrule

\ \ TwiBot-20\ \ & \ \ 229,573 \ \ &  \ \ \checkmark \ \  & \ \  \checkmark \ \  & \ \ \checkmark \ \  & \ \ 2020 \ \  \\

Cresci-17~\cite{cresci2017paradigm} & 9,813 & \checkmark & \checkmark & & 2017 \\

PAN-19\footnoteref{PAN-19} & 11,378 & \checkmark & & & 2019 \\
 
\bottomrule
\end{tabular}
\vspace{-10pt}
\end{table}

A summary of these three datasets is presented in Table~\ref{tab:dataset}. For three datasets, We randomly conduct a 7:2:1 partition as training, validation, and test set. Such a partition is shared across all experiments in Section ~\ref{subsec:bigexp}, ~\ref{subsec:sizeexp}, ~\ref{subsec:informationexp} and ~\ref{subsec:diversityexp}.

\noindent \textbf{Baselines.} We introduce competitive and state-of-the-art bot detection methods adopted in the experiments:

\begin{itemize} [leftmargin=*]
\item Lee \textit{et al.}~\cite{lee2011seven}: Lee \textit{et al.} use random forest classifier with feature engineering.

\item Yang \textit{et al.}~\cite{yang2020scalable}: Yang \textit{et al.} use random forest with user metadata and derived secondary features.

\item Kudugunta \textit{et al.}~\cite{kudugunta2018deep}: Kudugunta \textit{et al.} propose to use LSTM and fully connected neural networks to jointly leverage tweet semantics and user metadata.

\item Wei \textit{et al.}~\cite{wei2019twitter}: Wei \textit{et al.} use word embeddings and a three-layer BiLSTM to encode tweets and identify bots.

\item Miller \textit{et al.}~\cite{miller2014twitter}: Miller \textit{et al.} use 107 features and modified stream clustering algorithm for bot detection.

\item Cresci \textit{et al.}~\cite{cresci2016dna}: Cresci \textit{et al.} encodes user activities with strings and computes longest common substring to capture bot in groups.

\item Botometer~\cite{davis2016botornot}: Botometer 
is a publicly available demo that leverages more than one thousand features to identify Twitter bots on demand.

\item Alhosseini \textit{et al.}~\cite{ali2019detect}: Alhosseini \textit{et al.} adopt graph convolutional network for bot detection.
\end{itemize}

\noindent \textbf{Evaluation Metrics.}
We adopt Accuracy, F1-score and MCC~\cite{matthews1975comparison} as evaluation metrics. Accuracy is a straightforward metric while F1-score and MCC are more balanced alternatives.


\subsection{Benchmark performance}
\label{subsec:bigexp}

\begin{table*}
 \caption{The overall Twitter bot detection performance of various methods on our proposed TwiBot-20 and two public datasets, Cresci-17~\cite{cresci2017paradigm} and PAN-19$^3$. “/” denotes that the dataset doesn't have sufficient user information to support the method.}
 \label{tab:TwiBotMetric}
 \begin{tabular}{c c c c c c c c c c c c c c c c c c c c c c c c c c c c c c}
 \toprule
    \multicolumn{2}{c}{} &
    \multicolumn{3}{c}{\tabincell{c}{Lee \textit{et} \textit{al.}\\~\cite{lee2011seven}}}  &
    \multicolumn{3}{c}{\tabincell{c}{Yang \textit{et al.}\\~\cite{yang2020scalable}}} &
    \multicolumn{3}{c}{\tabincell{c}{Kudugunta \textit{et al.}\\~\cite{kudugunta2018deep}}} &
    \multicolumn{3}{c}{\tabincell{c}{Wei \textit{et} \textit{al.}\\~\cite{wei2019twitter}}} &
    \multicolumn{3}{c}{\tabincell{c}{Miller \textit{et al.}\\~\cite{miller2014twitter}}} &
    \multicolumn{3}{c}{\tabincell{c}{Cresci \textit{et al.}\\~\cite{cresci2016dna}}} &
    \multicolumn{3}{c}{\tabincell{c}{\tabincell{c}{Botometer \\~\cite{davis2016botornot}}}} &
    \multicolumn{3}{c}{\tabincell{c}{Alhosseini \textit{et al.}\\~\cite{ali2019detect}}} \\
   
 \midrule
 
 \multirow{3}{*}{\textbf{TwiBot-20}} &
 Acc & 
 \multicolumn{3}{c}{0.7456} &
 \multicolumn{3}{c}{\textbf{0.8191}} &
 \multicolumn{3}{c}{0.8174} &
 \multicolumn{3}{c}{0.7126} &
 \multicolumn{3}{c}{0.4801} &
 \multicolumn{3}{c}{0.4793} &
 \multicolumn{3}{c}{0.5584} & 
 \multicolumn{3}{c}{0.6813} \\
 
 &
 F1&
 \multicolumn{3}{c}{0.7823} &
 \multicolumn{3}{c}{\textbf{0.8546}} &
 \multicolumn{3}{c}{0.7517} &
 \multicolumn{3}{c}{0.7533} &
 \multicolumn{3}{c}{0.6266} &
 \multicolumn{3}{c}{0.1072} &
 \multicolumn{3}{c}{0.4892} &
 \multicolumn{3}{c}{0.7318} \\

 &
 MCC &
 \multicolumn{3}{c}{0.4879} &
 \multicolumn{3}{c}{0.6643} &
 \multicolumn{3}{c}{\textbf{0.6710}} &
 \multicolumn{3}{c}{0.4193} & 
 \multicolumn{3}{c}{-0.1372} & 
 \multicolumn{3}{c}{0.0839} & 
 \multicolumn{3}{c}{0.1558} & 
 \multicolumn{3}{c}{0.3543} \\
 
 \midrule
 \multirow{3}{*}{\textbf{Cresci-17}} & 
 Acc & 
 \multicolumn{3}{c}{0.9750} & 
 \multicolumn{3}{c}{\textbf{0.9847}} & 
 \multicolumn{3}{c}{0.9799} & 
 \multicolumn{3}{c}{0.9670} & 
 \multicolumn{3}{c}{0.5204} & 
 \multicolumn{3}{c}{0.4029} & 
 \multicolumn{3}{c}{0.9597} & 
 \multicolumn{3}{c}{/}  \\
 
 &
 F1&
 \multicolumn{3}{c}{0.9826} &
 \multicolumn{3}{c}{\textbf{0.9893}} &
 \multicolumn{3}{c}{0.9641} & 
 \multicolumn{3}{c}{0.9768} & 
 \multicolumn{3}{c}{0.4737} & 
 \multicolumn{3}{c}{0.2923} & 
 \multicolumn{3}{c}{0.9731} & 
 \multicolumn{3}{c}{/} \\

 &
 MCC &
 \multicolumn{3}{c}{0.9387} &
 \multicolumn{3}{c}{\textbf{0.9625}} &
 \multicolumn{3}{c}{0.9501} &
 \multicolumn{3}{c}{0.9200} &
 \multicolumn{3}{c}{0.1573} &
 \multicolumn{3}{c}{0.2255} &
 \multicolumn{3}{c}{0.8926} &
 \multicolumn{3}{c}{/} \\
 
 \midrule 
 \multirow{3}{*}{\textbf{PAN-19}\footnoteref{PAN-19}} &
 Acc &
 \multicolumn{3}{c}{/} &
 \multicolumn{3}{c}{/} &
 \multicolumn{3}{c}{/} &
 \multicolumn{3}{c}{\textbf{0.9464}} &
 \multicolumn{3}{c}{/} &
 \multicolumn{3}{c}{0.8797} &
 \multicolumn{3}{c}{/} &
 \multicolumn{3}{c}{/} \\
 
 &
 F1&
 \multicolumn{3}{c}{/} &
 \multicolumn{3}{c}{/} &
 \multicolumn{3}{c}{/} &
 \multicolumn{3}{c}{\textbf{0.9448}} &
 \multicolumn{3}{c}{/} &
 \multicolumn{3}{c}{0.8701} &
 \multicolumn{3}{c}{/} &
 \multicolumn{3}{c}{/} \\

 &
 MCC &
 \multicolumn{3}{c}{/} &
 \multicolumn{3}{c}{/} &
 \multicolumn{3}{c}{/} &
 \multicolumn{3}{c}{\textbf{0.8948}} &
 \multicolumn{3}{c}{/} &
 \multicolumn{3}{c}{0.7685} &
 \multicolumn{3}{c}{/} &
 \multicolumn{3}{c}{/} \\

 \bottomrule
\end{tabular}
\end{table*}

 Table ~\ref{tab:TwiBotMetric} reports bot detection performance of different methods on two public datasets cresci-17, PAN-19\footnoteref{PAN-19} and our proposed TwiBot-20. Table ~\ref{tab:TwiBotMetric} demonstrates that:
 \begin{itemize}[leftmargin=*]
     \item All bot detection baselines achieve significantly lower performance on TwiBot-20 than on cresci-17 or PAN-19\footnoteref{PAN-19}. This indicates that our TwiBot-20 is more challenging and social media bot detection is still an open problem.
     \item Alhosseini \textit{et al.}~\cite{ali2019detect} applies graph convolutional network to the task of Twitter bot detection, which demands that bot detection datasets include user neighborhood information. As Cresci \textit{et al.}~\cite{cresci2018reaction} points out, studying user communities is essential in future bot detection endeavors, where TwiBot-20 enables it by providing users' neighborhood information and the two other datasets fall short.
     \item From the comparison between Wei \textit{et al.}~\cite{wei2019twitter} and Kudugunta \textit{et al.}~\cite{kudugunta2018deep}, where both baselines use semantic information but the latter method also leverages user properties, it is demonstrated that the performance gap between two baselines is significantly larger on our proposed TwiBot-20 than on cresci-17. This indicates that apart from being a toy example, TwiBot-20 is relatively more complex, where bot detectors need to leverage more user information in order to perform well.
     \item Botometer~\cite{davis2016botornot} is a publicly available bot detection demo. Although it succeeds in capturing bots in cresci-17 where users in the dataset were collected back in 2017, it fails to match its previous performance on TwiBot-20, where users are collected in 2020. This demonstrates that the real-world Twittersphere has shifted and Twitter bots have evolved to evade previous detection methods, which calls for new research efforts and more up-to-date benchmarks like TwiBot-20.
     \item For feature engineering based methods such as Lee \textit{et al.}~\cite{lee2011seven} and Yang \textit{et al.}~\cite{yang2020scalable}, their performance drops significantly from cresci-17 to TwiBot-20. This trend indicates that failing to incorporate semantic and neighborhood information leads to worse performance on more recent datasets. This could again be attributed to the evolution of Twitter bots, thus future bot detection methods should leverage increasingly diversified and multi-modal user information to achieve desirable performance.
 \end{itemize}

\begin{table*}
 \caption{Modalities of Twitter user information that each compared method uses is summarized in this table. The definition of semantic, property and neighborhood user information follows the problem definition in Section ~\ref{sec:problemdefinition}.}
 \label{tab:BaselineSPN}
 \begin{tabular}{c c c c c c c c c} 
 \toprule
  &
  \tabincell{c}{Lee \textit{et} \textit{al.}\\~\cite{lee2011seven}} &
  \tabincell{c}{Yang \textit{et al.}\\~\cite{yang2020scalable}} &
  \tabincell{c}{Kudugunta \textit{et al.}\\~\cite{kudugunta2018deep}} &
  \tabincell{c}{Wei \textit{et} \textit{al.}\\~\cite{wei2019twitter}} &
  \tabincell{c}{Miller \textit{et al.}\\~\cite{miller2014twitter}} &
  \tabincell{c}{Cresci \textit{et al.}\\~\cite{cresci2016dna}} &
  \tabincell{c}{Botometer\\ ~\cite{davis2016botornot}} &
  \tabincell{c}{Alhosseini \textit{et al.}\\~\cite{ali2019detect}} 
  \\
  \midrule
   
  $\bf Semantic$ &
  \checkmark &  
  &  
  \checkmark &  
  \checkmark &  
  \checkmark &  
  \checkmark &  
  \checkmark &  
  \\
 
  $\bf Property$ &
  \checkmark &  
  \checkmark &  
  \checkmark &  
  &  
  \checkmark &  
  &  
  \checkmark &  
  \checkmark  \\
   
  $\bf Neighbor$ &
  &  
  &  
  &  
  &  
  &  
  &  
  \checkmark &  
  \checkmark  \\

 \bottomrule
\end{tabular}
\end{table*}
 
\begin{figure}[h]
  \centering
  \includegraphics[width=\linewidth]{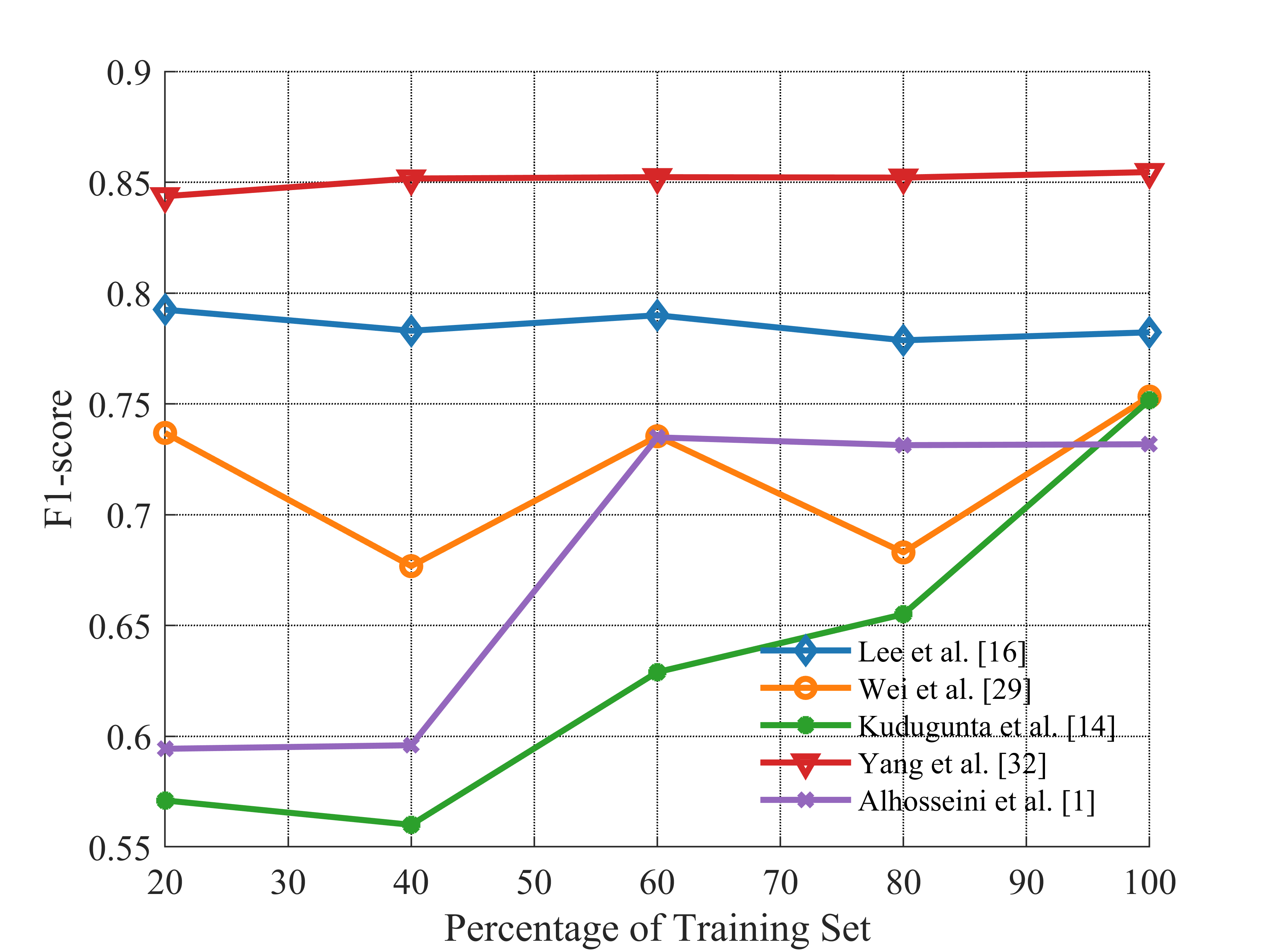}
  \caption{Baseline performance when trained on different percentage of the TwiBot-20 training set. Note that bot detection performance is evaluated on the full test set of TwiBot-20.}
  \Description{Limited Dataset Size}
  \label{fig:limited_size}
\end{figure}

\subsection{Dataset Size Study}
\label{subsec:sizeexp}
TwiBot-20 contains more Twitter users than any other known bot detection dataset and covers more types of accounts compared with existing bot detection benchmarks. In this section, we analyze the benefits of the enlarged scale and the challenge of more diversified users. We randomly choose different proportions of users from the Twibot-20 training set and compare the model performances trained with different data sizes in Figure ~\ref{fig:limited_size}.

Figure ~\ref{fig:limited_size} demonstrates that even trained on a small portion of our proposed TwiBot-20, competitive baselines like Yang \textit{et al.}~\cite{yang2020scalable} still maintain their performance. This indicates that TwiBot-20 can sufficiently train bot detectors, while other datasets with significantly fewer users would not stably benchmark bot detection methods.

\subsection{User Information Study}
\label{subsec:informationexp}
The task of Twitter bot detection is different from the standardized tasks in computer vision and natural language processing in that the input of bot detectors greatly vary from one another. Previous methods rely heavily on either tweet semantics analysis or user profile feature extraction, while methods that stress following behaviour and the graph structure it forms is on the rise. We summarize competitive bot detection baselines and their usage of multi-modal user information in Table \ref{tab:BaselineSPN}.

Along with the experiment results in Table ~\ref{tab:TwiBotMetric}, we make the following observations:

\begin{itemize}[leftmargin=*]
    \item To the best of our knowledge, TwiBot-20 is the first publicly available Twitter bot detection dataset to provide follow relationships between users to allow community-based detection measures. By providing multi-modal semantic, property and neighborhood user information, TwiBot-20 successfully supports all baseline bot detectors with varying demand of user information, while previous benchmarks fail to support newer research efforts such as Alhosseini \textit{et al.}~\cite{ali2019detect}.
    \item According to Table ~\ref{tab:BaselineSPN}, Kudugunta \textit{et al.}~\cite{kudugunta2018deep} leverages semantic and property information while Wei \textit{et al.}~\cite{wei2019twitter} only uses semantic information. It is demonstrated in Table ~\ref{tab:TwiBotMetric} that Kudugunta \textit{et al.}~\cite{kudugunta2018deep} outperforms Wei \textit{et al.}~\cite{wei2019twitter} on our proposed TwiBot-20. A similar contrast could be found between Alhosseini \textit{et al.}~\cite{ali2019detect} and Miller \textit{et al.}~\cite{miller2014twitter}. These performance gaps suggest that robust bot detection methods should leverage as much user information as possible.
\end{itemize}

Therefore, TwiBot-20 would be the ideal dataset to suffice the need for multi-modal user information and fairly evaluate any previous or future bot detectors.


\subsection{User Diversity Study}
\label{subsec:diversityexp}
\begin{figure}[h]
  \centering
  \includegraphics[width=\linewidth]{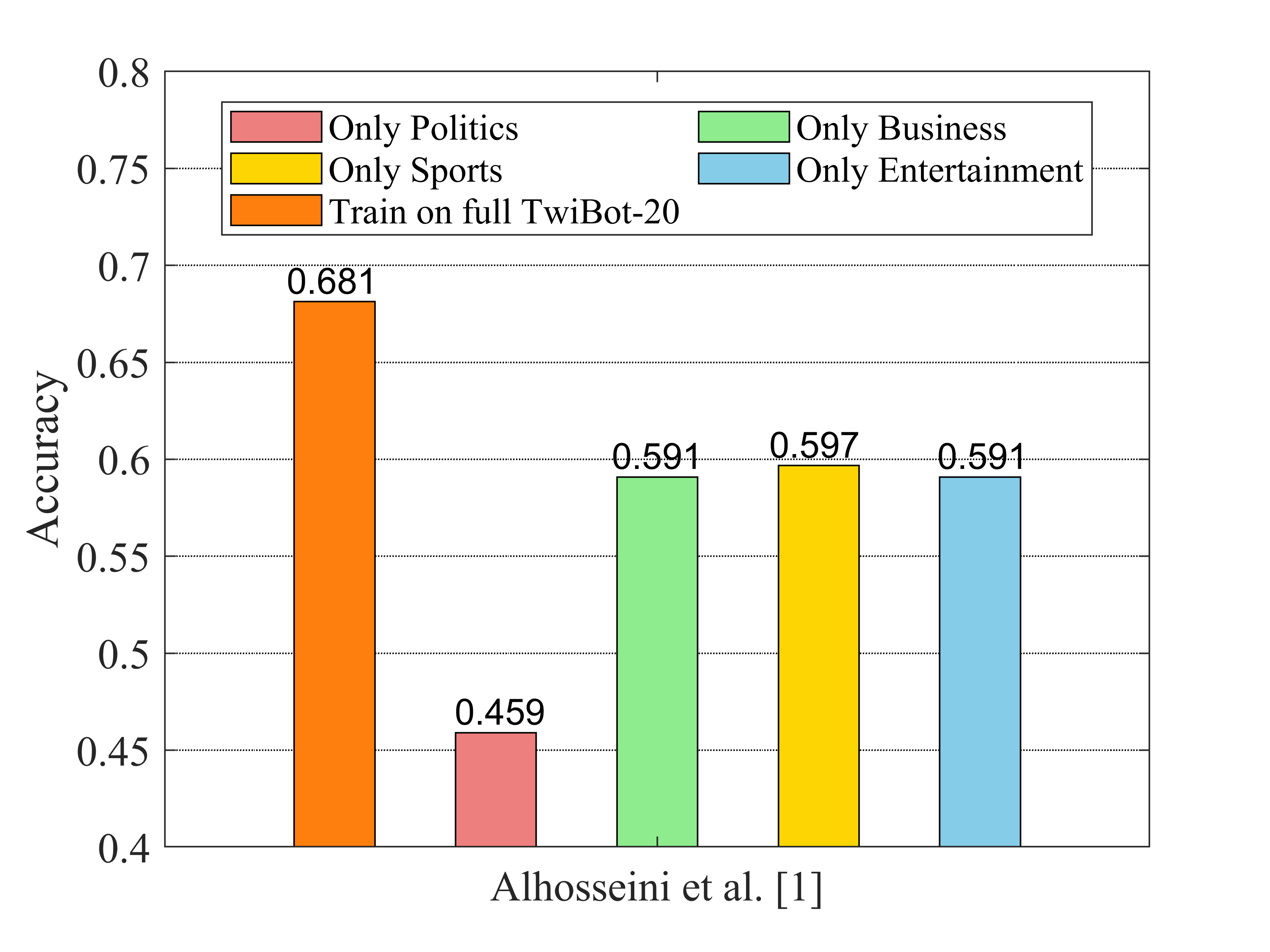}
  \caption{The community-based Alhosseini \textit{et al.}~\cite{ali2019detect}'s performance when trained on one interest domain in comparison to training on the full TwiBot-20. Note that bot detection performance is evaluated on the full test set of TwiBot-20.}
  \Description{user diversity}
  \label{fig:user_diversity}
\end{figure}

Previous datasets often focus on several specific types of Twitter bots and fall short of comprehensiveness. Another important aim of our proposed TwiBot-20 is to provide a stable benchmark that evaluates bot detectors' ability to identify diversified bots that co-exist on online social media. To prove that achieving good performance on one type of bot doesn't necessarily indicate the ability to identify diversified bots, we train the community-based bot detector Alhosseini \textit{et al.}~\cite{ali2019detect} on only one of the four interest domains in TwiBot-20 and evaluate it on the full test set. We present its performance in Figure  ~\ref{fig:user_diversity}.

Figure~\ref{fig:user_diversity} illustrates that when the method is trained on one of the user interest domains, Alhosseini \textit{et al.}~\cite{ali2019detect} fails to match its performance when it is trained on the full TwiBot-20. As a result, TwiBot-20 could better evaluate bot detection measures in that it contains diversified bots and genuine users, which demands bot detectors to jointly capture different types of bots rather than being limited to a specific bot detection scenario.

\section{Conclusion and Future Work}
\label{sec:conclusion}
Social media bot detection is attracting increasing research interests in recent years. We collected and annotated Twitter data to present a comprehensive Twitter bot detection benchmark TwiBot-20, which is representative of the diversified Twittersphere and captures different types of bots that co-exist on major social media platforms. We make TwiBot-20 public, hoping that it would alleviate the lack of comprehensive datasets in Twitter bot detection and facilitate further research. Extensive experiments demonstrate that state-of-the-art bot detectors fail to match their previously reported performance on TwiBot-20, which shows that Twitter bot detection is still a challenging task and demands continual efforts. In the future, we plan to study novel Twitter bots and propose robust bot detectors.


\bibliographystyle{ACM-Reference-Format}
\bibliography{sample-base}

\end{document}